\documentclass[prl,twocolumn,aps,superscriptaddress]{revtex4}
\usepackage{amsfonts}

\usepackage{dcolumn}
\usepackage{bm}
\usepackage{tikz}
\usepackage{graphicx}
\usepackage[colorlinks=true,citecolor=blue,urlcolor=blue]{hyperref}
\usepackage{amsmath,amsfonts,amssymb,times,natbib}
\usepackage[standard]{ntheorem}
\usepackage{CJK}

\begin{document}


\title{Direct measurement of density matrix elements using phase-shifting technique}
\author{Tianfeng Feng}

\affiliation{State Key Laboratory of Optoelectronic Materials and Technologies and School of Physics, Sun Yat-sen University, Guangzhou, People's Republic of China}

\author{Changliang Ren}
\email{renchangliang@hunnu.edu.cn}
\affiliation{Key Laboratory of Low-Dimensional Quantum Structures and Quantum Control of Ministry of Education, Key Laboratory for Matter Microstructure and Function of Hunan Province, Department of Physics and Synergetic Innovation Center for Quantum Effects and Applications, Hunan Normal University, Changsha 410081, China}

\author{Xiaoqi Zhou}
\email{zhouxq8@mail.sysu.edu.cn}
\affiliation{State Key Laboratory of Optoelectronic Materials and Technologies and School of Physics, Sun Yat-sen University, Guangzhou, People's Republic of China}

\date{\today}
\begin{abstract}
Direct measurement protocol allows reconstructing specific elements of the density matrix of a quantum state without using quantum state tomography. However, the direct measurement protocols to date are primarily based on weak or strong measurements with ancillary pointer, which interacts with the investigated system to extract information about the specified elements. Here we present a new direct measurement scheme based on phase-shifting operations which do not need ancillary pointers. In this method, estimates of at most 6 expectation values of
projective observables suffice
 to determine any specific element of an unknown quantum density matrix. 
A concrete quantum circuit to implement this direct measurement protocol for multi-qubit states is provided, which is composed of just single-qubit gates and two multi-qubit controlled-phase gates. This scheme is also extended for direct measurement of the density matrix of continuous-variable quantum states. 
Our method can be used in quantum information applications where only partial information about the quantum state needs to be extracted, for example, problems such as entanglement witnessing, fidelity estimation of quantum systems, and quantum coherence estimation.
\end{abstract}


\maketitle
\section{I. Introduction}
The density matrix of quantum states is at the heart of quantum mechanical description of nature and crucial for understanding and quantifying quantum phenomena.  Characterizing the density matrix of quantum states is thus a fundamental task in various quantum applications. The standard way of measuring the quantum density matrix is the so-called quantum state tomography (QST) \cite{Vogel,Smithey,White,James}, which is an established method of characterizing quantum systems that involves performing a complete set of measurements on  a series of bases followed by data processing to infer the density matrix. As the dimension of the Hilbert space of the investigated quantum system increases, the number of measurements for QST and the complexity of the reconstruction algorithm increases accordingly. Substantial efforts have been made to reduce the number of measurements for QST using the methods such as compressed sensing \cite{Compressed,Compressed2} and symmetric informationally complete positive operator-valued measures \cite{B}. However, for many quantum applications \cite{entanglement,coherence,Verification,Pauli,Pauli2,Huang}, such as the detection of quantum entanglement \cite{entanglement} and quantum coherence \cite{coherence}, and the verification of quantum states \cite{Verification}, it is not necessary to reconstruct the full density matrix but only a fraction of it. In such cases, QST becomes inefficient and finding a way to directly measure specific elements of a density matrix becomes crucial.

\begin{figure}
\includegraphics[width=0.33\textwidth]{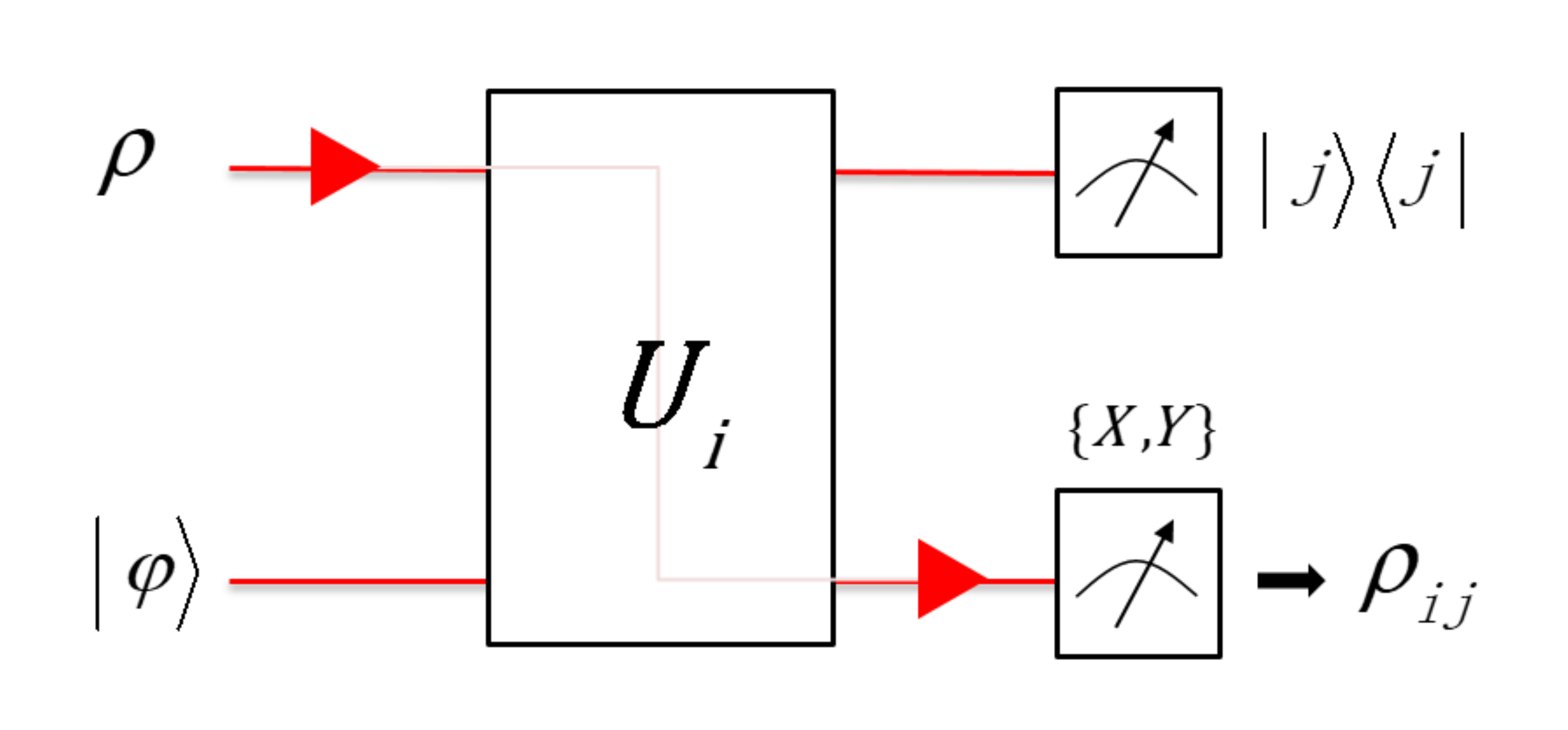}
\caption{ Direct measurement of the density matrix of a general quantum state using weak measurement on a pointer. The arrow denotes the information flow from the system to the pointer. One may directly reconstruct the non-diagonal term $\rho_{ij}$ from the results of the measurements on the pointer. }
\label{fig1}
\end{figure}

Recently, a new method called direct measurement (DM) has been developed to directly measure specific elements of a density matrix with the help of weak measurements \cite{Lundeen,Lundeen2}. The procedure of the DM scheme can be summarized as follows (see Fig. \ref{fig1}):
The information about the element $\rho_{ij}$ of the density matrix of quantum systems may be transferred to an ancillary pointer by the joint evolution $U_i$, along with a postselection of quantum state $|j\rangle \langle j|$. Then, by measuring the pointer state, one may directly reconstruct the desired element $\rho_{ij}$.
Nowadays, much attentions has been paid to DM technique. In particular, it has been shown that DM may be implemented using strong measurements\cite{Vallone,Calderaro}, and an optimal DM scheme has been proposed\cite{Ren}, in which only a single pointer is required. Besides, various experimental demonstrations have been presented~\cite{Thekkadath,Salvail,Malik,Zhang,Pan,CDM,DMNC,DMP}, such as characterizing $27$-dimensional quantum state \cite{Malik} and direct measurement of a nonlocal entangled two-photon states \cite{Pan}.

\textcolor{black}{Although DM does not require complicated reconstruction techniques to determinate specific elements of density matrix, the presence of the ancillary pointers themselves adds complexity to the measurement procedure and sometimes a suitable pointer is not easy to be constructed \cite{Haapasalo,Maccone}}. 
\textcolor{black}{Some studies have begun to explore the state measurement scheme without auxiliary system, such as $\delta$-quench protocol \cite{delta}, but so far they have been limited to very specific systems.} In this paper, 
we provide a general framework using phase-shifting technique for directly measuring the density matrix of generic quantum states without ancillary pointers, which can be seen as an extension  of $\delta$-quench protocol \cite{delta}. In our method, estimates of at most 6 expectation values of projective observables suffice to determine any specific element of an unknown density matrix. We present a quantum circuit that implements the phase-shifting measurement protocol in the multi-qubit case, involving only single-qubit gates and two multi-qubit controlled-phase gates. In addition, we also show that the protocol can be extended to the continuous-variable quantum systems.

\section{II.   Measuring the elements of a density matrix}
 In general, the density matrix of a $d$-level quantum system can be expressed as $\rho=\sum^{d-1}_{i,j=0}\rho_{ij}|i\rangle\langle j|$, where $|i\rangle$ denotes the elements of a complete orthonormal basis. To fully characterize the density matrix, one should determine at least $d^2-1$ real parameters. Determination of the diagonal element $\rho_{ii}$ of the density matrix is a trivial task since one can measure these elements on the computational basis, while extracting the information of the non-diagonal element $\rho_{ij}$ ($i\ne j$) is a challenging task. 
 
 One way to get the non-diagonal element $\rho_{ij}$ is to measure the operator $M_{j,i}=|j\rangle\langle i|$, where $\rho_{ij}=\mathrm{Tr}(\rho M_{j,i})$. However, $M_{j,i}$ is a non-Hermitian operator, and its measurement requires the use of the interaction with ancillary systems. For example, in the traditional DM schemes \cite{Lundeen2, Calderaro}, two sequential weak measurements on $|i\rangle\langle i|$ and $|+\rangle\langle +|$, which are realized by using ancillary systems, followed by a projective measurement on $|j\rangle\langle j|$ are required to implement the measurement of $M_{j,i}$, where $|+\rangle=\frac{1}{\sqrt{d}}\sum_{i=0}^{d-1}|i\rangle$ \cite{Lundeen}.

Unlike the traditional DM methods, $\rho_{ij}$ can also be obtained by measuring $h$ different Hermitian operators $\hat{K}_{i,j}^{(g)}$, where
 \begin{equation}
 \rho_{ij}=\sum_{g=1}^{h} \eta_g \mathrm{Tr}(\rho \hat{K}_{i,j}^{(g)}),
 \label{eq1}
 \end{equation}
in which $\eta_g$ is the coefficient associated with $\hat{K}_{i,j}^{(g)}$. 
The decomposition of $\rho_{ij}$ given in Eq. (1) is not unique, and there exist many different ways of decomposition. The criterion for a good decomposition should be that the number of measurement operators is as small as possible while satisfying the condition that $\hat{K}_{i,j}^{(g)}$ is easy for measurement.

\begin{figure}
\includegraphics[width=0.49\textwidth]{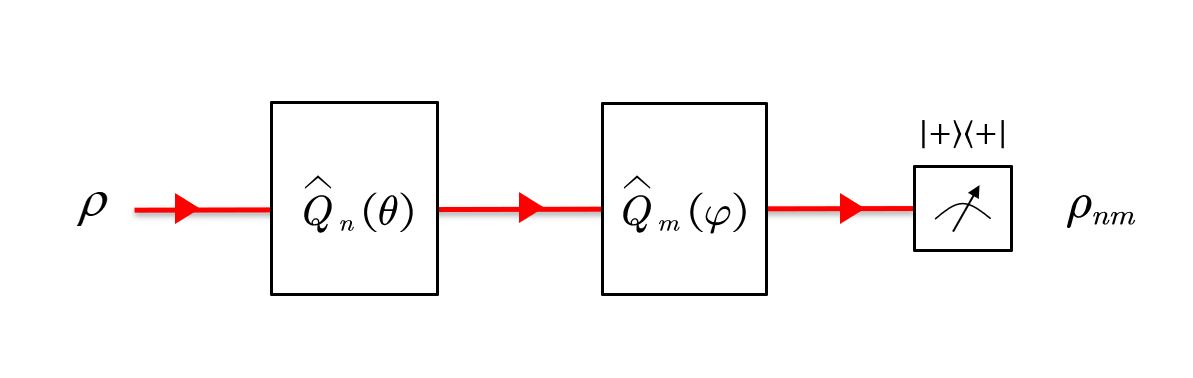}
\caption{\textcolor{black}{Phase-shifting measurement protocol.  The general quantum circuit of phase-shifting measurement. An unknown quantum state $\rho$ can be directly measured via a projector $\hat{K}_{n,m}^{(\theta,\varphi)}$ consisting of two phase-shifting operator $\hat{Q}_n(\theta)$ and $\hat{Q}_m(\varphi)$ along with a  post-selection $|+\rangle \langle+|$, where $\theta=\{0,\frac{\pi}{2}, -\frac{\pi}{2}\}$ and $\varphi=\{0,\pi\}$.}}
\label{figps}
\end{figure}

\section{iii. Phase-shifting technique for direct measurement of density matrix elements}

Here we present a scheme using phase-shifting operations \cite{Calderaro,delta,Ren2} which is able to calculate $\rho_{ij}$ ($i \ne j$) by measuring only 6 projective operators. The projective operators we want to measure are of the following form
 \begin{equation}
 \hat{K}_{n,m}^{(\theta,\varphi)}=\hat{Q}^{\dagger}_n(\theta)\hat{Q}^{\dagger}_m(\varphi)| +\rangle \langle +|\hat{Q}_m(\varphi)\hat{Q}_n(\theta),
 \end{equation}
where $\hat{Q}_n(\theta)$ and $\hat{Q}_m(\varphi)$ are phase-shifting operators. $\hat{Q}_n(\theta)$ operator is used to add a relative phase $e^{i\theta}$ to the component $|n\rangle$ and is defined as
\begin{equation}
 \hat{Q}_n(\theta)=\hat{I}+(e^{i\theta}-1)|n\rangle \langle n|,
 \end{equation}
where $0\le n\le d-1$ and $\theta\in[0, 2\pi]$. As shown in Fig. 2, the measurement of the operator $\hat{K}_{n,m}^{(\theta,\varphi)}$ on the quantum state $\rho$ is achieved by successively performing the phase-shifting operations $\hat{Q}_n(\theta)$ and $\hat{Q}_m(\varphi)$ on $\rho$ and then projecting it onto $| +\rangle$. 
From the measurement results, the expected value of $\hat{K}_{n,m}^{(\theta,\varphi)}$ can be obtained as
\begin{equation}
\begin{split}
\langle \hat{K}_{n,m}^{(\theta,\varphi)}\rangle&=\mathrm{Tr}[\rho \hat{K}_{n,m}^{(\theta,\varphi)}]\\
&=\langle +|\hat{Q}_m(\varphi)\hat{Q}_n(\theta)\rho \hat{Q}^{\dagger}_n(\theta) \hat{Q}^{\dagger}_m(\varphi)| +\rangle.
\label{eq2}
\end{split}
\end{equation}
Substituting Eqn.~(3) into Eqn. (4), one can obtain
\begin{equation}
\begin{aligned}
\langle \hat{K}_{n,m}^{(\theta,\varphi)}\rangle &= 
 \frac{2}{d}[\mathrm{cos}(\theta-\varphi)-\mathrm{cos}\varphi]Re[\rho_{nm}]\\ &
-\frac{2}{d}[\mathrm{sin}(\theta-\varphi)-\mathrm{sin}\varphi]Im[\rho_{nm}]\\&
+p_m(\varphi)+s_{n,m}(\theta), 
\end{aligned}
\end{equation}
where  $p_m(\varphi)=\langle + |\hat{Q}_m(\varphi) \rho \hat{Q}^{\dagger}_m(\varphi)|+\rangle$ and $s_{n,m}(\theta)=\frac{2}{d}(1-\mathrm{cos}\theta)\langle n|\rho|n\rangle+\frac{1}{d}(e^{i\theta}-1)\sum_{i=0,i\ne m}^{d-1}\langle n|\rho|i\rangle+\frac{1}{d}(e^{-i\theta}-1)\sum_{i=0,i\ne m}^{d-1}\langle i|\rho|n\rangle$. By choosing the values of $(\theta, \varphi)$ as $(0, 0)$, $(0, \pi)$, $(\frac{\pi}{2}, 0)$, $(\frac{\pi}{2}, \pi)$,  $(-\frac{\pi}{2}, 0)$ and $(-\frac{\pi}{2}, \pi)$, respectively, one can get 6 expected values. Using these 6 expectation values, $p_m(\varphi)$ and $s_{n,m}(\theta)$ can be cancelled out to calculate the real and imaginary parts of $\rho_{nm}$ (See Appendix A)
\begin{equation}
\begin{aligned}
\mathrm{Re}{[\rho_{nm}]} &=\frac{d}{8}[2(\langle \hat{K}_{n,m}^{(0,0)}\rangle-\langle \hat{K}_{n,m}^{(0,\pi)}\rangle) - \\&
(\langle \hat{K}_{n,m}^{(\frac{\pi}{2},0)}\rangle-\langle \hat{K}_{n,m}^{(\frac{\pi}{2},\pi)}\rangle)-(\langle \hat{K}_{n,m}^{(-\frac{\pi}{2},0)}\rangle -\langle \hat{K}_{n,m}^{(-\frac{\pi}{2},\pi)}\rangle)],\\
 \mathrm{Im}{[\rho_{nm}]} &= -\frac{d}{8}[(\langle \hat{K}_{n,m}^{(\frac{\pi}{2},0)}\rangle-\langle \hat{K}_{n,m}^{(\frac{\pi}{2},\pi)}\rangle)\\&
-(\langle \hat{K}_{n,m}^{(-\frac{\pi}{2},0)}\rangle -\langle \hat{K}_{n,m}^{(-\frac{\pi}{2},\pi)}\rangle)].&
\end{aligned}
\end{equation}

\begin{figure}
\includegraphics[width=0.5\textwidth]{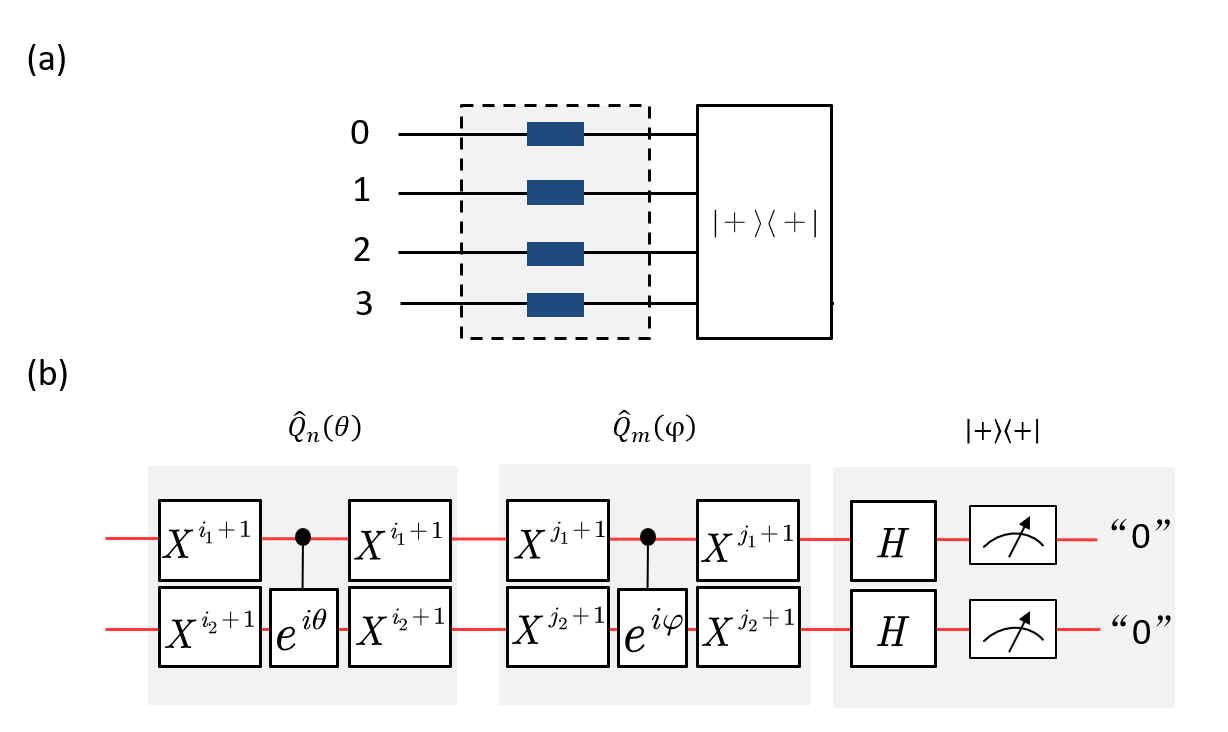}
\caption{\textcolor{black}{The quantum circuit of phase-shifting measurements on a two-qubit system. (a)  Phase-shifting measurement of a single photon state. One can see that each blue square represents a phase shifter, which introduces a relative phase in its corresponding mode. The phase-shifting operator $\hat{Q}_n(\theta)$ and $\hat{Q}_m(\varphi)$ can be implemented by two phase shifters in the mode $n$ and $m$, which add relative phase $e^{i\theta}$ and $e^{i\varphi}$ respectively. (b) The phase-shifting operator $\hat{Q}_n(\theta)$ is decomposed to four single-qubit NOT gates and a two-qubit controlled-phase gate. The post-selection is realized by a Hadamard gate with a measurement outcome $00$ in computational basis. }}
\label{figps}
\end{figure}

\section{IV.  Implementation of phase-shifting technique in multi-qubit systems}

For a single high-dimensional quantum state, our scheme is easy to implement, and we illustrate it below with a single photon state in a multi-spatial mode.
As shown in Fig. 3a, there is a phase shifter (blue box) on each spatial mode of this photon, and it is only necessary to load $\theta$ in the $n$th phase shifter and $\varphi$  in the $m$th phase shifter to realize the phase-shifting operators of $\hat{Q}_n(\theta)$ and $\hat{Q}_m(\varphi)$, and then just project the quantum state to $|+\rangle$ to complete the operation.



For a multi-qubit quantum system, our scheme requires the use of controlled-phase gates, and in the following we illustrate the scheme with a two-qubit quantum system as an example.
As shown in Fig. 3b, the phase-shifting operator $\hat{Q}_n(\theta)$  is decomposed into four single-qubit $X$ gates and a two-qubit controlled-phase gate, which can be efficiently realized in various quantum systems \cite{fSim1,fSim2}. 
The first two single-qubit gates $X^{i_1+1}$ and $X^{i_2+1}$ are used to convert $|i_1\rangle|i_2\rangle$ to $|1\rangle|1\rangle$, where $i_1 i_2$ is the binary form of $n$. The two-qubit controlled-phase gate is used to load the $|1\rangle|1\rangle$ component with the phase $e^{i\theta}$. The next two single-qubit gates $X^{i_1+1}$ and $X^{i_2+1}$ are used to convert $|1\rangle|1\rangle$ back to $|i_1\rangle|i_2\rangle$. In total, these four single-qubit gates and the two-qubit controlled-phase gate realize the phase-shifting operator $\hat{Q}_n(\theta)$, namely adding the phase $e^{i\theta}$ to the $n$th component. Similarly, $\hat{Q}_m(\varphi)$ can be implemented in the same way. After that, the two-qubit quantum state is projected onto $|+\rangle$, which is achieved by two single-qubit Hadamard gates and two single-qubit projective operations onto $|0\rangle$. Note that our approach can be easily extended to the case of multi-qubit quantum states (See Appendix B).

\section{V. Phase-shifting technique for continuous-variable quantum systems.} 

In addition to being applied to quantum systems with discrete variables, our scheme can also be applied to quantum systems with continuous variables. For a continuous-variable quantum system, the phase-shifting operator becomes \cite{delta}
\begin{equation}
 \hat{Q}_{x^{'}}(\theta)=\hat{I}+\int d x\delta(x-x^{'})(e^{i\theta}-1)|x\rangle \langle x|,
 \end{equation}
 where $\{ |x\rangle \}$ corresponds to a set of orthogonal bases for the quantum system.
The expectation value of the projective operator $\hat{K}_{x^{'},x^{''}}^{(\theta,\theta^{'})}$ is now
\begin{equation}
	\langle \hat{K}_{x^{'},x^{''}}^{(\theta,\varphi)} \rangle=\langle +|\hat{Q}_{x^{''}}(\varphi)\hat{Q}_{x^{'}}(\theta)\rho \hat{Q}^{\dagger}_{x^{'}}(\theta)\hat{Q}^{\dagger}_{x^{''}}(\varphi)| +\rangle,
\end{equation}
where $|+\rangle=\int|x\rangle dx$. 
Similarly, with six expectation values, one can calculate any specific element of a continuous-variable quantum state's density matrix (See Appendix C). Here we note that, in practical experiments, the results of measurements performed on continuous variable systems are indeed discrete because of the limited accuracy of the measurement instruments. Our approach can also be generalized to directly reconstruct the Wigner function \cite{Wigner32}. In addition, we propose a feasible experimental scheme for the direct measurement of the continuous-time quantum density matrix (See Appendix D).

\section{VI.  Conclusion $\&$ Outlook}


In summary, we have presented a general framework for the measurement of density matrices using phase-shifting techniques. Unlike conventional direct measurement schemes, our scheme allows the direct measurement of quantum density matrix elements without the aid of ancillary quantum systems. It is applicable to quantum systems of discrete as well as continuous variables, whether it is a pure or mixed states. 

	 Our method can measure specific elements of the density matrix, and it has the potential to be used in applications like fidelity estimation of quantum systems \cite{fidelity} where only a fraction of the elements of the density matrix need to be obtained. 
For example, it is sufficient to evaluate the fidelity of an N-qubit quantum state with respect to an N-qubit GHZ state by just measuring four elements of the density matrix of this quantum state, which is because, although the density matrix of an N-qubit GHZ states is $2^n\times 2^n$-dimensional, it has only four elements in the corners that are non-zero terms. For similar reasons, our method is also expected to be applied to other quantum information applications including entanglement witnessing \cite{entanglement} and quantum coherence estimation \cite{coherence}. 

In our scheme, the quantum state is post-selected at the state $|+\rangle$, which results in all other components orthogonal to $|+\rangle$ being discarded. As an outlook, it is an interesting direction of investigation whether these discarded components can also be used for density matrix element measurements.

\begin{acknowledgments}
This work was supported by     National Key R$\&$D Program of China (2017YFA0305200), The Key R$\&$D Program of Guangdong Province (2018B030329001, 2018B030325001), The National Natural Science Foundation of China (61974168, 12075245), the Natural Science Foundation of Hunan Province (2021JJ10033), and Xiaoxiang Scholars Programme of Hunan Normal university.
\end{acknowledgments}


\begin{thebibliography}{}
	
	
	
	\bibitem{Vogel}
	K. Vogel and H. Risken, Determination of quasiprobability distributions in terms of probability distributions for the rotated quadrature phase, Phys. Rev. A \textbf{40}, 2847 (1989).
	\bibitem{Smithey}
	D. T. Smithey,  M. Beck, M. G. Raymer, and A. Faridani, Measurement of the Wigner distribution and the density matrix of a light mode using optical homodyne tomography: application to squeezed states and the vacuum, Phys. Rev. Lett. \textbf{70}, 1244 (1993).
	\bibitem{White}
	A. G. White, D. F. V. James, P. H. Eberhard, and P. G. Kwiat, Nonmaximally Entangled States: Production, Characterization, and Utilization, Phys. Rev. Lett. \textbf{83}, 3103 (1999).
	\bibitem{James}
	D. F. V. James, P. G. Kwiat, W. J. Munro, and A. G. White, Measurement of qubits, Phys. Rev. A \textbf{64}, 052312 (2001).
	
	
	\bibitem{Compressed}
	D. Gross, Y.-K. Liu, S. T. Flammia, S. Becker, and J. Eisert, Quantum State Tomography via Compressed Sensing, Phys. Rev. Lett. \textbf{105}, 150401 (2010).
	\bibitem{Compressed2}
	A. Kalev, R. L. Kosut, and I. H. Deutsch, Quantum tomog- raphy protocols with positivity are compressed sensing protocols, npj Quantum Inf. \textbf{1}, 15018 (2015).
	
	
	
	\bibitem{B} N. Bent, H. Qassim, A. A. Tahir, D. Sych, G. Leuchs, L. L. Sanchez-Soto, E. Karimi, and R. W. Boyd, Experimental Realization of Quantum Tomography of Photonic Qudits via Symmetric Informationally Complete Positive Operator-Valued Measures, Phys. Rev. X \textbf{5}, 041006 (2015).
	
	
	\bibitem{entanglement}
	R. Horodecki, P. Horodecki, M. Horodecki, and K. Horodecki, Quantum entanglement, Rev. Mod. Phys. \textbf{81}, 865 (2009).
	\bibitem{coherence}
	H. Xu, F. Xu, T. Theurer, D. Egloff, Z.-W. Liu, N. Yu, M. B. Plenio, and L. Zhang, Experimental Quantification of Coherence of a Tunable Quantum Detector, Phys. Rev. Lett.\textbf{125}, 060404 (2020)
	\bibitem{Verification}
	S. Pallister, N. Linden, and A. Montanaro, Optimal Verification of Entangled States with Local Measurements, Phys. Rev. Lett. \textbf{120}, 170502 (2018)
	\bibitem{Pauli}
	S. T. Flammia1 and Y.-K. Liu, Direct Fidelity Estimation from Few Pauli Measurements, Phys. Rev. Lett. \textbf{106}, 230501 (2011)
	\bibitem{Pauli2}
	M. P. da Silva, O. Landon-Cardinal, and D. Poulin, Practical Characterization of Quantum Devices without Tomography, Phys. Rev. Lett. \textbf{107}, 210404 (2011)
	\bibitem{Huang}
	H.-Y. Huang, R. Kueng, and J. Preskill, Predicting many properties of a quantum system from very few measurements, Nat. Physics \textbf{16}, 1050-1057 (2020)
	
	
	
	\bibitem{Lundeen}
	J. S. Lundeen, B. Sutherland, A. Patel, C. Stewart, and C. Bamber, Direct measurement of the quantum wave function, Nature (London) \textbf{474}, 188 (2011).
	\bibitem{Lundeen2}
	J. S. Lundeen and C. Bamber, Procedure for Direct Measurement of General Quantum States Using Weak Measurement, Phys. Rev. Lett. \textbf{108}, 070402 (2012).
	
	
	\bibitem{Vallone}
	G. Vallone and D. Dequal, Strong Measurements give a Better Direct Measurement of the Quantum Wave Function, Phys. Rev. Lett. \textbf{116}, 040502 (2016).
	
	\bibitem{Calderaro}
	L. Calderaro, G. Foletto, D. Dequal, P. Villoresi, and G. Vallone, Direct Reconstruction of the Quantum Density Matrix by Strong Measurements, Phys. Rev. Lett. \textbf{121}, 230501 (2018).
	
	
	
	\bibitem{Ren}
	C. Ren, Y. Wang, and J. Du, Efficient Direct Measurement of Arbitrary Quantum Systems via Weak Measurement, Phys. Rev. Appl. \textbf{123}, 150402 (2019).
	
	\bibitem{Malik}
	M. Malik, M. Mirhosseini, M. P. J. Lavery, J. Leach, M. J. Padgett, and R.W. Boyd, Direct measurement of a 27-dimensional orbital-angular-momentum state vector, Nat. Commun. \textbf{5}, 3115 (2014).
	
	\bibitem{Thekkadath}
	G. S. Thekkadath, L. Giner, Y. Chalich, M. J. Horton, J. Banker, and J. S. Lundeen, Direct Measurement of the Density Matrix of a Quantum System, Phys. Rev. Lett. \textbf{117}, 120401 (2016).
	\bibitem{Zhang}
	C. Zhang, M. Hu, Z. Hou, J. Tang, J. Zhu, G. Xiang, C. F. Li, G. C. Guo, and Y. S. Zhang, Direct Measurement of the Two-Dimensional Spatial Quantum Wavefunction via Strong Measurements, arxiv:1811.01560.
	\bibitem{Salvail}
	J. Z. Salvail, M. Agnew, A. S. Johnson, E. Bolduc, J. Leach, and R.W. Boyd, Full characterization of polarization states of light via direct measurement, Nat. Photonics \textbf{7}, 316 (2013).
	\bibitem{Pan}
	W. Pan, X. Xu, Y. Kedem, Q. Wang, Z. Chen, M. Jan, K. Sun, J. Xu, Y. Han, C. Li, and G. Guo, Direct Measurement of a Nonlocal Entangled Quantum State, Phys. Rev. Lett. \textbf{123}, 150402 (2019).
	\bibitem{CDM}
	M. Mirhosseini, O. S. Magana-Loaiza, S. M. H. Rafsanjani, and R. W. Boyd, Compressive Direct Measurement of the Quantum Wave Function, Phys. Rev. Lett. \textbf{113}, 090402 (2014)
	\bibitem{DMNC}
	E. Bolduc, G. Gariepy, adn J. Leach, Direct measurement of large-scale quantum states via expectation values of non-Hermitian matrices, Nat. Commun. \textbf{7}, 10439 (2016)
	\bibitem{DMP}
	Y. Kim1, Y.-S. Kim, S.-Y. Lee, S.-W. Han, S. Moon, Y.-H. Kim, and Y.-W. Cho, Direct quantum process tomography via measuring sequential weak values of incompatible observables, Nat. Commun. \textbf{9}, 192 (2018)
	
	
	\bibitem{Haapasalo}
	E. Haapasalo, P. Lahti, and J. Schultz, Weak versus approximate values in quantum state determination, Phys. Rev. A \textbf{84}, 052107 (2011)
	\bibitem{Maccone}
	L. Maccone and C. C. Rusconi, State estimation: A comparison between direct state measurement and tomography, Phys. Rev. A \textbf{89}, 022122 (2014).
	
	
	\bibitem{delta}
	S. Zhang, Y. Zhou, Y. Mei, K. Liao, Y. Wen \emph{et al}, $\delta$-Quench Measurement of a Pure Quantum-State Wave Function, Phys. Rev. Lett. \textbf{123}, 190402 (2019)
	
	\bibitem{Ren2}
	H. F. Hofmann and C. Ren, Direct observation of temporal coherence by weak projective measurements of photon arrival time, Phys. Rev. A \textbf{87}, 062109 (2013).
	

	
	
	
	
	
	\bibitem{fSim1}
	I. D. Kivlichan, J. McClean, N. Wiebe, C. Gidney, A. Aspuru-Guzik, G. K. Chan, and R. Babbush, Quantum simulation of electronic structure with linear depth and connectivity, Phys. Rev. Lett. \textbf{120}, 110501 (2018).
	
	\bibitem{fSim2}
	B. Foxen, C. Neill, A. Dunsworth, P. Roushan, B. Chiaro \emph{et al}, Demonstrating a Continuous Set of Two-qubit Gates for Near-term Quantum Algorithms, Phys. Rev. Lett. \textbf{125}, 120504 (2020)
	
	
	
	
	\bibitem{Wigner32}
	E. Wigner, On the quantum correction for thermodynamic equilibrium, Physical Review, \textbf{40}, 749 (1932).
	\bibitem{fidelity}
	Flammia, S. T.  and Liu, Y.-K. Direct fidelity estimation from few Pauli measurements. Phys. Rev. Lett. \textbf{106}, 230501 (2011).
	
	
	
	
	
	
	








\bibitem{Yang}
C. Yang, Z. Gu, P. Chen, Z. Qin, J. F. Chen, and W. Zhang, Tomography of the Temporal-Spectral State of Subnatural- Linewidth Single Photons from Atomic Ensembles, Phys. Rev. Applied \textbf{10}, 054011 (2018).
\bibitem{Wasilewski}
W. Wasilewski, P. Kolenderski, and R. Frankowski, Spectral Density Matrix of a Single Photon Measured, Phys. Rev. Lett. \textbf{99}, 123601 (2007).
\bibitem{Beduini}
F. A. Beduini, J. A. Zieliska, V. G. Lucivero, Y. A. de Icaza Astiz, and M.W. Mitchell, Interferometric Measurement of the Biphoton Wave Function, Phys. Rev. Lett. \textbf{113}, 183602 (2014).
\bibitem{Chen}
P. Chen, C. Shu, X. Guo, M. M. T. Loy, and S. Du, Measuring the Biphoton Temporal Wave Function with Polarization-Dependent and Time-Resolved Two-Photon Interference, Phys. Rev. Lett. \textbf{114}, 010401 (2015).
\bibitem{Davis}
A. O. C. Davis, V. Thiel, M. Karpiski, and B. J. Smith, Measuring the Single-Photon Temporal-Spectral Wave Function, Phys. Rev. Lett. \textbf{121}, 083602 (2018).
\bibitem{Ansari}
V. Ansari, J. M. Donohue, M. Allgaier, L. Sansoni, B. Brecht, J. Roslund, N. Treps, G. Harder, and C. Silberhorn, Tomography and Purification of the Temporal-Mode Structure of Quantum Light, Phys. Rev. Lett. \textbf{120}, 213601 (2018).
\bibitem{Davis2}
A. O. C. Davis, Valerian Thiel, Michal Karpinski, and Brian J. Smith, Measuring the Single-Photon Temporal-Spectral Wave Function, Phys. Rev. Lett. \textbf{121}, 083602 (2018).







\end{thebibliography}

\section{appendix A. Phase-shifting technique for quantum density matrix.}
In the main text, the expectation value $ \hat{K}_{n,m}^{(\theta,\theta^{'})}$ ($n\ne m$) is given as

\begin{equation}
\begin{aligned}
\langle \hat{K}_{n,m}^{(\theta,\varphi)}\rangle &=\langle +|\hat{Q}_m(\varphi)\hat{Q}_n(\theta)\rho \hat{Q}^{\dagger}_n(\theta) \hat{Q}^{\dagger}_m(\varphi)| +\rangle \\
&=
p_m(\varphi)+\frac{1}{\sqrt{d}}(e^{i\theta}-1)\langle n|\rho\hat{Q}^{\dagger}_m(\varphi)|+\rangle\\
&+\frac{1}{\sqrt{d}}(e^{-i\theta}-1)\langle +|\hat{Q}_m(\varphi)\rho|n\rangle\\
&+\frac{2}{d}(1-\mathrm{cos}\theta)\langle n|\rho|n\rangle,
\end{aligned}
\end{equation}
where  $p_m(\varphi)=\langle + |\hat{Q}_m(\varphi) \rho \hat{Q}^{\dagger}_m(\varphi)|+\rangle$. 
Because $\langle n| \rho|m\rangle=\rho_{nm}=\mathrm{Re}[\rho_{nm}]+i\mathrm{Im}[\rho_{nm}]$ and $\hat{Q}^{\dagger}_m(\varphi)|+\rangle=\sum_{i=0,i\ne m}^{d-1}\frac{1}{\sqrt{d}}|i\rangle+e^{-i\varphi}\frac{1}{\sqrt{d}}|m\rangle$, upon substituting them into above equation, one can rewrite $\langle \hat{K}_{n,m}^{(\theta,\varphi)}\rangle $ as
 
\begin{equation}
\begin{aligned}
\langle \hat{K}_{n,m}^{(\theta,\varphi)}\rangle &=
\frac{2}{d}(\mathrm{cos}(\theta-\varphi)-\mathrm{cos}\varphi)\mathrm{Re}[\rho_{nm}]\\
&-\frac{2}{d}(\mathrm{sin}(\theta-\varphi)-\mathrm{sin}\varphi)\mathrm{Im}[\rho_{nm}]\\
&+p_m(\varphi)+s_{n,m}(\theta),
\end{aligned}
\end{equation}
 where $s_{n,m}(\theta)=\frac{2}{d}(1-\mathrm{cos}\theta)\langle n|\rho|n\rangle+\frac{1}{d}(e^{i\theta}-1)\sum_{i=0,i\ne m}^{d-1}\langle n|\rho|i\rangle+\frac{1}{d}(e^{-i\theta}-1)\sum_{i=0,i\ne m}^{d-1}\langle i|\rho|n\rangle$. Setting $\varphi=0$ and $\pi$, we have 

\begin{equation}
\begin{aligned}
\langle \hat{K}_{n,m}^{(\theta,0}\rangle &=\frac{2}{d}(\mathrm{cos}\theta-1)\mathrm{Re}[\rho_{nm}]\\
&-\frac{2}{d}(\mathrm{sin}\theta)\mathrm{Im}[\rho_nm]+p_m(0)+s_{n,m}(\theta), \\
\langle \hat{K}_{n,m}^{(\theta,\pi}\rangle &=\frac{2}{d}(-\mathrm{cos}\theta+1)\mathrm{Re}[\rho_{nm}]\\
&-\frac{2}{d}(-\mathrm{sin}\theta)\mathrm{Im}[\rho_nm]+p_m(\pi)+s_{n,m}(\theta).
\end{aligned}
\end{equation}
Here we define a subtraction factor $\langle \hat{K}_{n,m}^{(\theta,0}\rangle -\langle \hat{K}_{n,m}^{(\theta,\pi}\rangle $, we obtain 
\begin{equation}
\begin{aligned}
\langle \hat{K}_{n,m}^{(\theta,0}\rangle -\langle \hat{K}_{n,m}^{(\theta,\pi}\rangle&=\frac{4}{d}(\mathrm{cos}\theta-1)\mathrm{Re}[\rho_{nm}]\\
&-\frac{4}{d}\mathrm{sin}\theta \mathrm{Im}[\rho_{nm}]+p_m(0)-p_m(\pi).
\end{aligned}
\end{equation}
Then we have
\begin{equation}
\begin{aligned}
\langle \hat{K}_{n,m}^{0,0}\rangle -\langle \hat{K}_{n,m}^{(0,\pi}\rangle&=p_m(0)-p_m(\pi),\\
\langle \hat{K}_{n,m}^{(\frac{\pi}{2},0}\rangle -\langle \hat{K}_{n,m}^{\frac{\pi}{2},\pi}\rangle&=-\frac{4}{d}\mathrm{Re}[\rho_{nm}]-\frac{4}{d}\mathrm{Im}[\rho_{nm}]\\
&+p_m(0)-p_m(\pi),\\
\langle \hat{K}_{n,m}^{(-\frac{\pi}{2},0}\rangle -\langle \hat{K}_{n,m}^{-\frac{\pi}{2},\pi}\rangle&=-\frac{4}{d}\mathrm{Re}[\rho_{nm}]+\frac{4}{d} \mathrm{Im}[\rho_{nm}]\\
&+p_m(0)-p_m(\pi).
\end{aligned}
\end{equation}
The expressions for the real and imaginary parts of non-diagonal element $\rho_{n,m}$ are as follows 
\begin{eqnarray}
	\begin{aligned}
\mathrm{Re}{[\rho_{nm}]} &=\frac{d}{8}[2(\langle \hat{K}_{n,m}^{(0,0)}\rangle-\langle \hat{K}_{n,m}^{(0,\pi)}\rangle)-\\
&(\langle \hat{K}_{n,m}^{(\frac{\pi}{2},0)}\rangle-\langle \hat{K}_{n,m}^{(\frac{\pi}{2},\pi)}\rangle)-(\langle \hat{K}_{n,m}^{(-\frac{\pi}{2},0)}\rangle -\langle \hat{K}_{n,m}^{(-\frac{\pi}{2},\pi)}\rangle)],\\
\mathrm{Im}{[\rho_{nm}]} &=-\frac{d}{8}[(\langle \hat{K}_{n,m}^{(\frac{\pi}{2},0)}\rangle-\langle \hat{K}_{n,m}^{(\frac{\pi}{2},\pi)}\rangle)-\\
&-(\langle \hat{K}_{n,m}^{(-\frac{\pi}{2},0)}\rangle -\langle \hat{K}_{n,m}^{(-\frac{\pi}{2},\pi)}\rangle)].
	\end{aligned}
\end{eqnarray}

\section{appendix B. The quantum circuit of direct measurement for multi-qubit systems.} 
\begin{figure}
\includegraphics[width=0.5\textwidth]{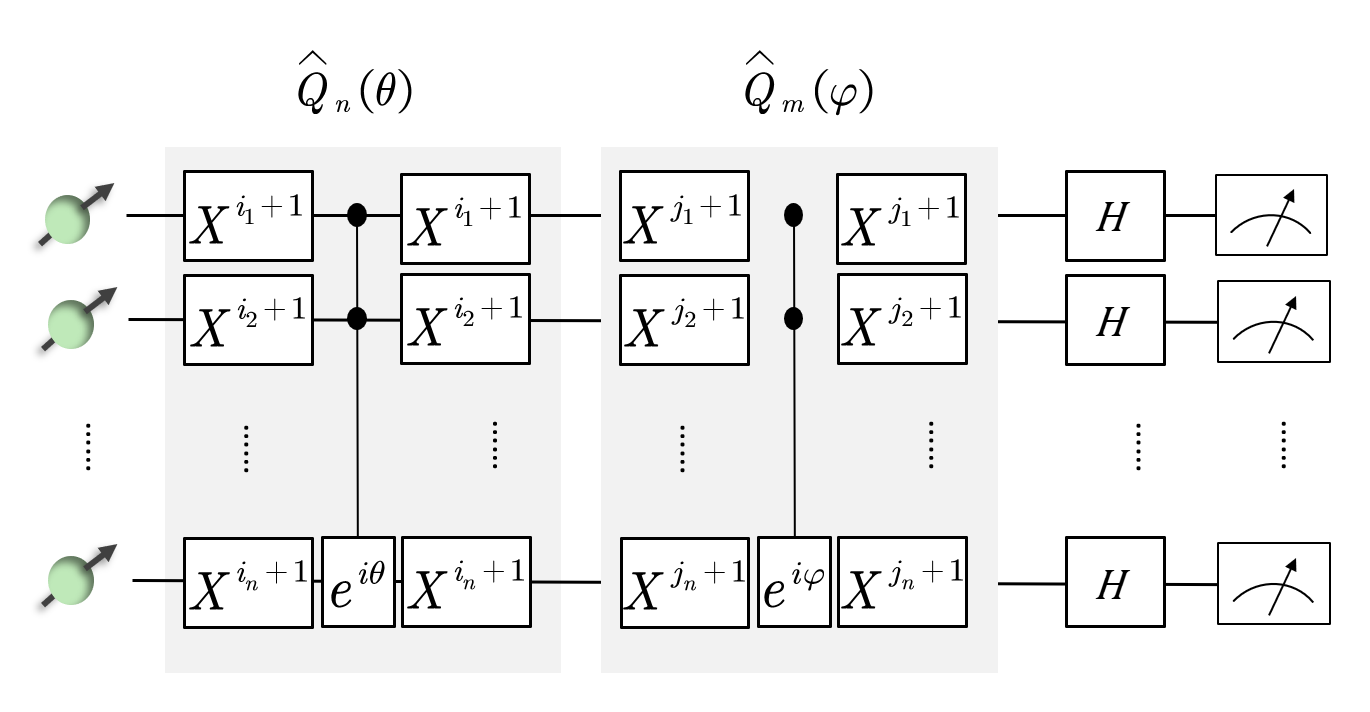}
\caption{ The quantum circuit of phase-shifting measurements for multi-qubit states. }
\label{circuit}
\end{figure}

Here we show how to extend to the multi-qubit case of phase-shifting protocol illustrated for the two-qubit case. 
As shown in Fig. \ref{circuit}, a general $N$-qubit state can be measured via the multi-qubit phase-shifting circuit, where $\hat{Q}_n(\theta)$ and $\hat{Q}_m(\varphi)$ are phase-shifting gates. The $\hat{Q}_n(\theta)$ can be decomposed into $2N$ single-qubit X gates in parallel and one multi-qubit phase gate with phase $\theta$. Similar to the two-qubit case, the single-qubit $X$ gates are used to scan the desired components. Therefore, the phase-shifting gate may be rewritten as
   \begin{equation}
  \hat{Q}_n(\theta)= I+(e^{i\theta}-1)\bigotimes_{k=1}^{N} X^{i_k+1}|1\rangle^{\otimes N}\langle1|^{\otimes N}\bigotimes_{k=1}^{N} X^{i_k+1},
   \end{equation}
 where $n=i_1i_2...i_{n}$ in binary~($i_k\in\{0,1\}$) and $X^{i_k+1}$ is the single-qubit $X$ gate acting on the $k$\emph{th} qubit.
In general, our method can be extended to the mulit-qudit~($d$-dimension) systems when the $X$ and $H$ gates are replaced by their corresponding $d$-dimensional gates, e.g. $X^{i_k+d-1}$.  

\section{appendix C. Phase-shifting technique for continuous-variable quantum systems.}
 In continuous Hilbert space, the phase-shifting operator
 $\hat{Q}_{x^{'}}(\theta)$ is given as
\begin{equation}
 \hat{Q}_{x^{'}}(\theta)=\hat{I}+\int d x\delta(x-x^{'})(e^{i\theta}-1)|x\rangle \langle x|.
 \end{equation}

After the first phase-shifting operator, the shifted state can be expressed as
\begin{equation}
\begin{aligned}
 \rho' &=\hat{Q}_{x^{'}}(\theta)\rho \hat{Q}^{\dagger}_{x^{'}}(\theta)\\
 &=\rho+(e^{i\theta}-1)|x^{'}\rangle \langle x^{'}|\rho+(e^{-i\theta}-1)\rho|x^{'}\rangle \langle x^{'}|\\
 &+(2-2\mathrm{cos}\theta)|x^{'}\rangle \langle x^{'}|\langle x^{'}|\rho|x^{'}\rangle.
\end{aligned}
\end{equation}

Without loss of generality, we suppose that $|+\rangle=\int|x\rangle dx$. Similar to the case of discrete variable, we define $|y_{x^{''}}(\theta{'})\rangle=\hat{Q}^{\dagger}_{x^{''}}(\varphi)|+\rangle$, as
\begin{equation}
|y_{x^{''}}(\varphi)\rangle=\int|x\rangle dx + \int dx \delta(x-x^{''})(e^{i\varphi}-1)|x\rangle \langle x|.
\end{equation}
Since the probability of postselection for the phase-shifted state can be calculated as
\begin{equation}
 \langle \hat{K}_{x^{'},x^{''}}^{(\theta,\varphi)} \rangle=\langle y_{x^{''}}(\theta{'})|\rho^{'}| y_{x^{''}}(\varphi)\rangle,
 \end{equation}
one can expand it, as
\begin{equation}
\begin{aligned}
 \langle \hat{K}_{x^{'},x^{''}}^{(\theta,\varphi)} \rangle&=\langle y_{x^{''}}(\varphi)|\rho|y_{x^{''}}(\theta{'})\rangle +(e^{i\theta}-1)\langle x^{'}|\rho|y_{x^{''}}(\varphi)\rangle\\
&+(e^{-i\theta}-1)\langle y_{x^{''}}(\theta{'})|\rho|x^{'}\rangle+2(1-\mathrm{cos}\theta)\langle x^{'}|\rho|x^{'}\rangle, &
 \end{aligned}
  \end{equation}
where the coherent information of density operator is contained in $\langle x^{'}|\rho|y_{x^{''}}(\theta{'})\rangle$, as
 \begin{equation}
 \langle x^{'}|\rho|y_{x^{''}}(\varphi)\rangle=\int_{x\neq x^{''}} dx \langle x^{'}|\rho|x\rangle + e^{-i\varphi}\langle x^{'}|\rho|x^{''}\rangle.
 \end{equation}

By setting $\varphi=\{0,\pi\}$, we have
\begin{equation}
\langle x^{'}|\rho|y_{x^{''}}(0)\rangle-\langle x^{'}|\rho|y_{x^{''}}(\pi)\rangle= 2\langle x^{'}|\rho|x^{''}\rangle.
\end{equation}
Upon substituting the above equation into Eqn.~(20) and set $\theta=\{\frac{\pi}{2},-\frac{\pi}{2}\}$, we obtain
\begin{eqnarray}
	\begin{aligned}
p_{x^{'},x^{''}}&=\langle \hat{K}_{x^{'},x^{''}}^{(\frac{\pi}{2},0)} \rangle -\langle \hat{K}_{x^{'},x^{''}}^{(\frac{\pi}{2},\pi)} \rangle\\
&=\langle +|\rho|+\rangle -\langle y_{x^{''}}(\pi)|\rho|y_{x^{''}}(\pi)\\
&-4(\mathrm{Re}{[\rho_{x^{'},x^{''}}]}+\mathrm{Im}{[\rho_{x^{'},x^{''}}]}),\\
p'_{x^{'},x^{''}}&=\langle \hat{K}_{x^{'},x^{''}}^{(-\frac{\pi}{2},0)} \rangle-\langle \hat{K}_{x^{'},x^{''}}^{(\frac{-\pi}{2},\pi)} \rangle\\
&=\langle +|\rho|+\rangle -\langle y_{x^{''}}(\pi)|\rho|y_{x^{''}}(\pi)\rangle\\
&-4(\mathrm{Re}{[\rho_{x^{'},x^{''}}]}-\mathrm{Im}{[\rho_{x^{'},x^{''}}]}).
\end{aligned}
\end{eqnarray}
Furthermore, we can solve the above two equations. Let $q=\langle +|\rho|+\rangle-\langle y_{x^{''}}(\pi)|\rho|y_{x^{''}}(\pi)\rangle$, the real and the imaginary parts  of $\rho_{x^{'},x^{''}}$ can be written as
\begin{eqnarray}
\mathrm{Re}{[\rho_{x^{'},x^{''}}]} =\frac{1}{8}(2q-p'_{x^{'},x^{''}}-p'_{x^{'},x^{''}}),\\
\mathrm{Im}{[\rho_{x^{'},x^{''}}]} =-\frac{1}{8}(p'_{x^{'},x^{''}}-p_{x^{'},x^{''}}).
\end{eqnarray}


\section{appendix D. Experimental proposal for measuring the density matrix of a temporal quantum state.} 
\begin{figure}
	\includegraphics[width=0.5\textwidth]{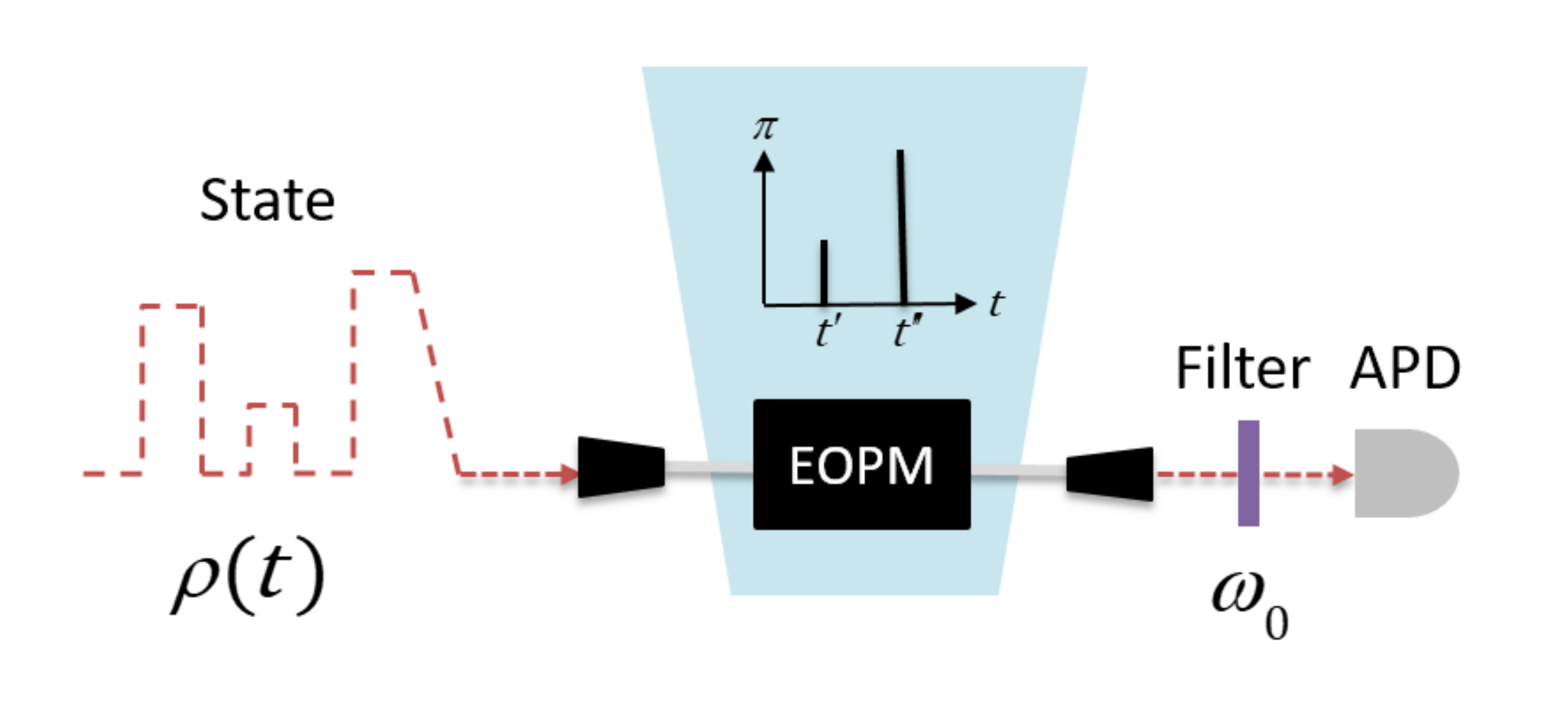}
	\caption{Experimental proposal for measuring the density operator of a temporal wavefunction or density matrix using the sequential phase-shifting protocol. An arbitrary temporal quantum state of photons can be measured by double phase-shifting operation by EOPM and postselected by a filter with a center frequency at $\omega_0$. }
	\label{proposal}
\end{figure}

We now show that our method is an ideal method for measuring the temporal quantum states, or equivalently the density matrix $\rho(t)$.
So far, much attention has been paid to measuring the temporal quantum wave function of single-photon and two-photon\cite{delta,Yang,Wasilewski,Beduini,Chen,Davis,Ansari,Davis2}.
Here we consider a feasible experimental scheme to measure a temporal quantum states (or density matrix $\rho(t)$) by using our scheme and the experimental setup is presented in \cite{delta}. As shown in Fig. \ref{proposal},  $\rho(t)$ is phase-shifted by an electro-optic phase modulator~(EOPM) with phase $\theta=\{0, \frac{\pi}{2},-\frac{\pi}{2}\}$ and $\varphi=\{0, \pi\}$ at $t^{'}$ and $t^{''}$, respectively. The filter, an optical cavity with resonance angular frequency $\omega_0$, performs post-selection projection while a fast avalanche photon detector~(APD) is used to detect the photons.
The whole process only requires a double phase-shifting operation in the quantum state and one postselection.

{}

\end{document}